\documentclass{cernyrep}
\usepackage[bookmarks, colorlinks=true, linktoc=page, pdftex, linkcolor=black, citecolor=black, urlcolor=blue]{hyperref}
\usepackage{graphbox}
\usepackage{fancyhdr}
\fancyhfoffset{4 mm}
\fancypagestyle{ARTTITLE}{%
\fancyhf{} 
\lhead{\small{Proceedings of the 2018 CERN--Accelerator--School course on \it{Beam Instrumentation}, Tuusula, (Finland)}} 
\lfoot{Available online at \url{https://cas.web.cern.ch/previous-schools}}
\rfoot{\thepage\hspace*{3mm}}
 
}

\begin{document}
\title{Transverse emittance}
\author{E. Bravin}
\institute{CERN, Geneva, Switzerland}

\begin{abstract}
This chapter defines the concept of transverse emittance and describes the techniques most frequently used to measure it.
\end{abstract}

\keywords{CAS, School, particle accelerator; phase space, transverse emittance; beam profile; beam size.}

\maketitle
\thispagestyle{ARTTITLE}

\section{Introduction}
The knowledge and the prediction of the distribution of the charged particles inside a particle accelerator is of fundamental importance for its design and operation.
In other lectures at this school, the theory describing the dynamics of a charged particle in a magnetic lattice has been explained. A particle beam contains, however, a large number of particles making it impossible to study them individually. Statistical quantities that summarize the status of the beam as a whole are needed.

The transverse emittance, as we will see, is an invariant quantity, i.e. a quantity that is conserved along the magnetic structures of an accelerator complex, that, together with the optics parameters of the beam line, describes the distribution of the particles in the transverse phase space.

The transverse emittance is a statistical quantity that can be defined in different ways. Usually accelerator scientists use different definitions for different cases, either to simplify the study or to emphasize certain aspects. The measurement of the transverse emittance can also be accomplished using different techniques, each one better suited for certain situations. It is important to understand the difference between the definitions and the approximations introduced by each measurement method.

The measurement of the transverse emittance relies heavily on the techniques described in the transverse profiles lecture. Often the emittance is calculated by sampling the transverse profile at one or more locations.
In particular, it is important to understand the distribution of the particles in the geometric space and in the phase space and the way the two spaces are related.

The sampling of particles happens almost exclusively in the geometric space as measuring the direction of each particle independently is impossible. We then use models to infer the distribution in phase space starting from the distribution is geometrical space. It is like inferring the shape and size of objects from the shadows they project.

\section{Definition of transverse emittance}
\begin{figure}[htb]
\centering
\includegraphics[width=0.4\linewidth]{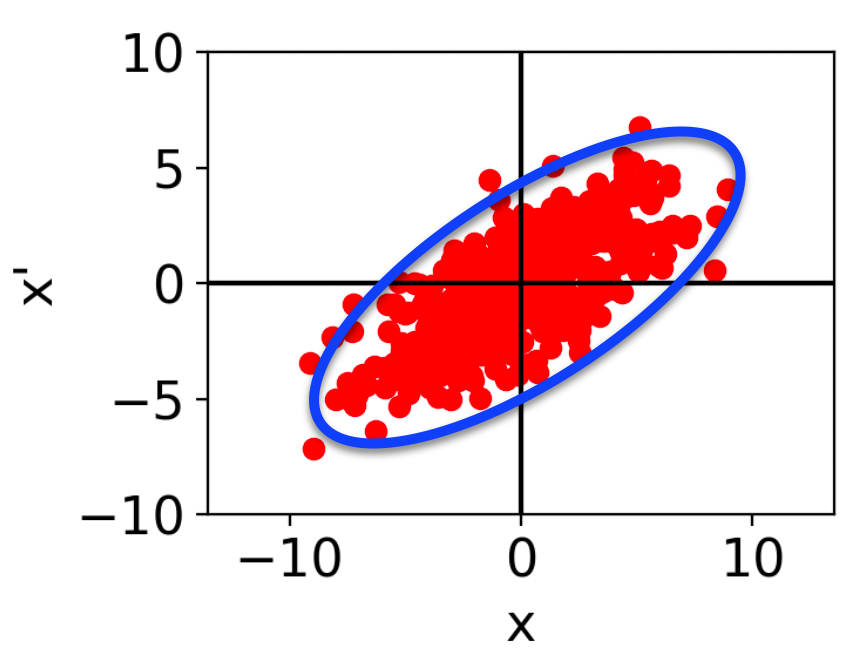}
\includegraphics[width=0.4\linewidth]{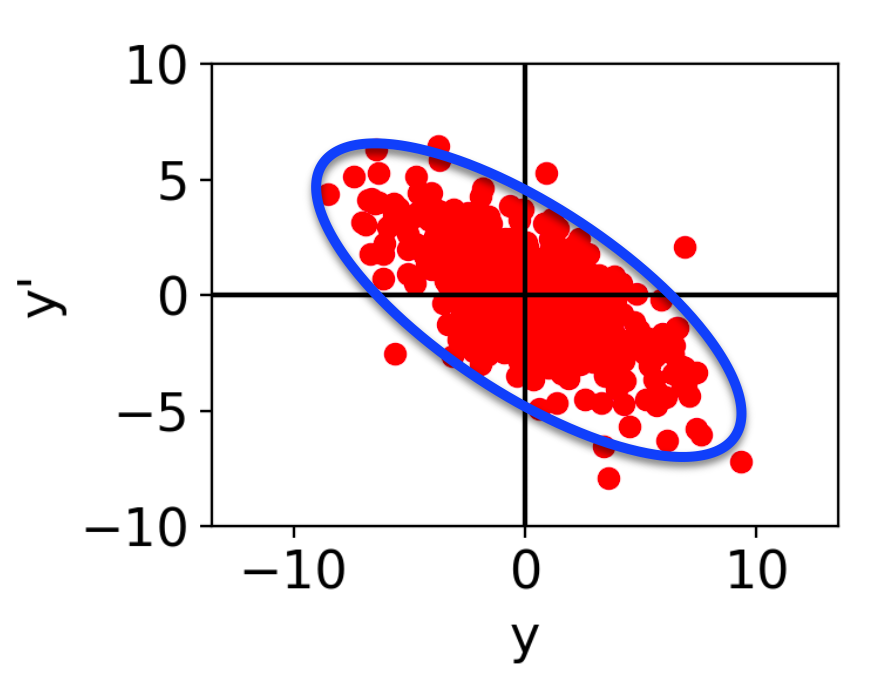}
\caption{Particle distribution in the horizontal (left) and vertical (right) transverse phase spaces. The blue ellipses represent the smallest ellipses containing 95\% of the particles.}
\label{fig:phasespaces}
\end{figure}

In \Fref{fig:phasespaces} typical phase space distributions are shown, ideally these are bivariate normal distributions, but the real cases can differ substantially from it.
No matter what the distribution is it is possible to define few interesting quantities like the first and second momenta of the distributions. We can define a statistical emittance, called \emph{rms-emittance}, using these quantities~\cite{Flottmann:2003pw}. In particular we use the root-mean-square of $x$ and $x'$ (or $y$ and $y'$) plus the correlation-product-terms.

\begin{figure}[htb]
\centering
\includegraphics[align=c, width=0.4\linewidth]{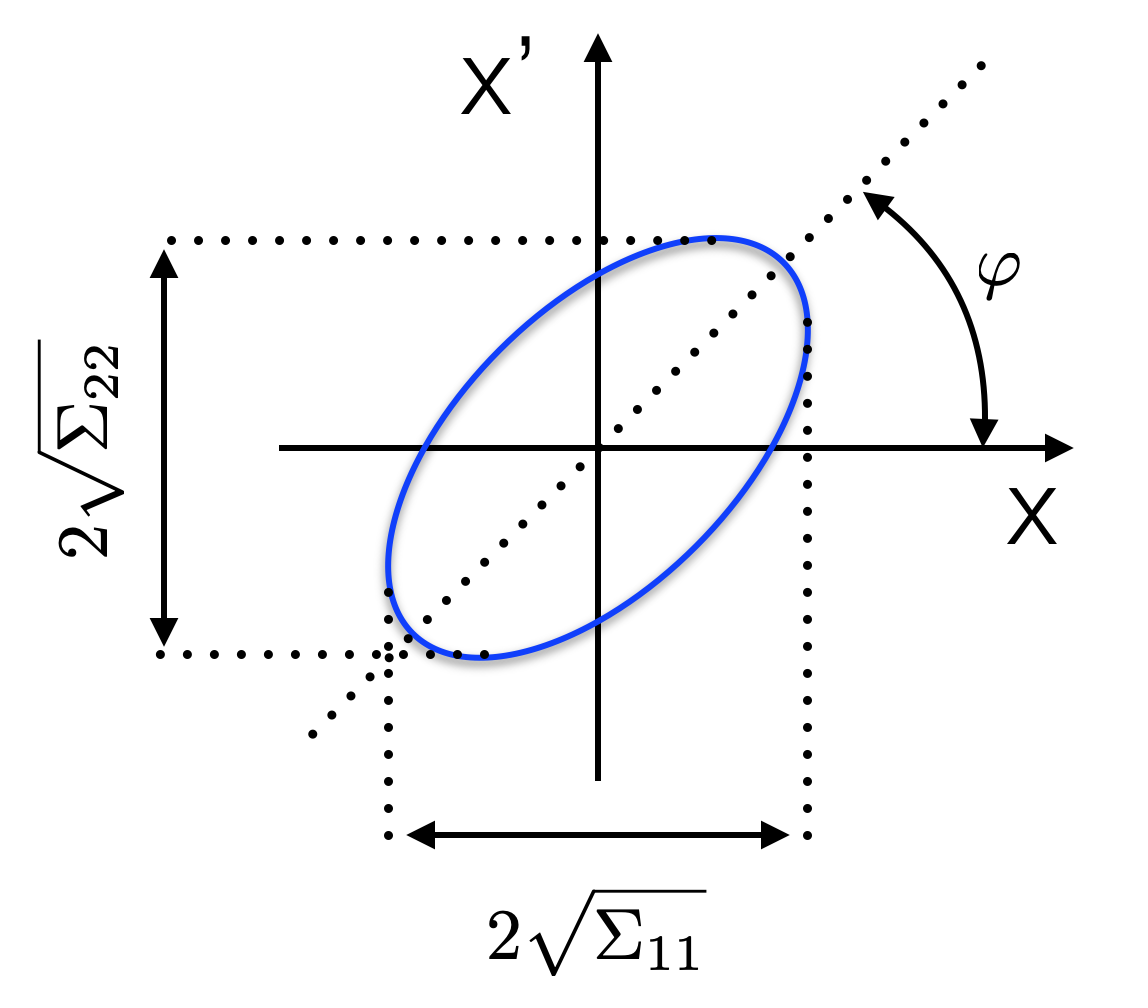}
\includegraphics[align=c, width=0.3\linewidth]{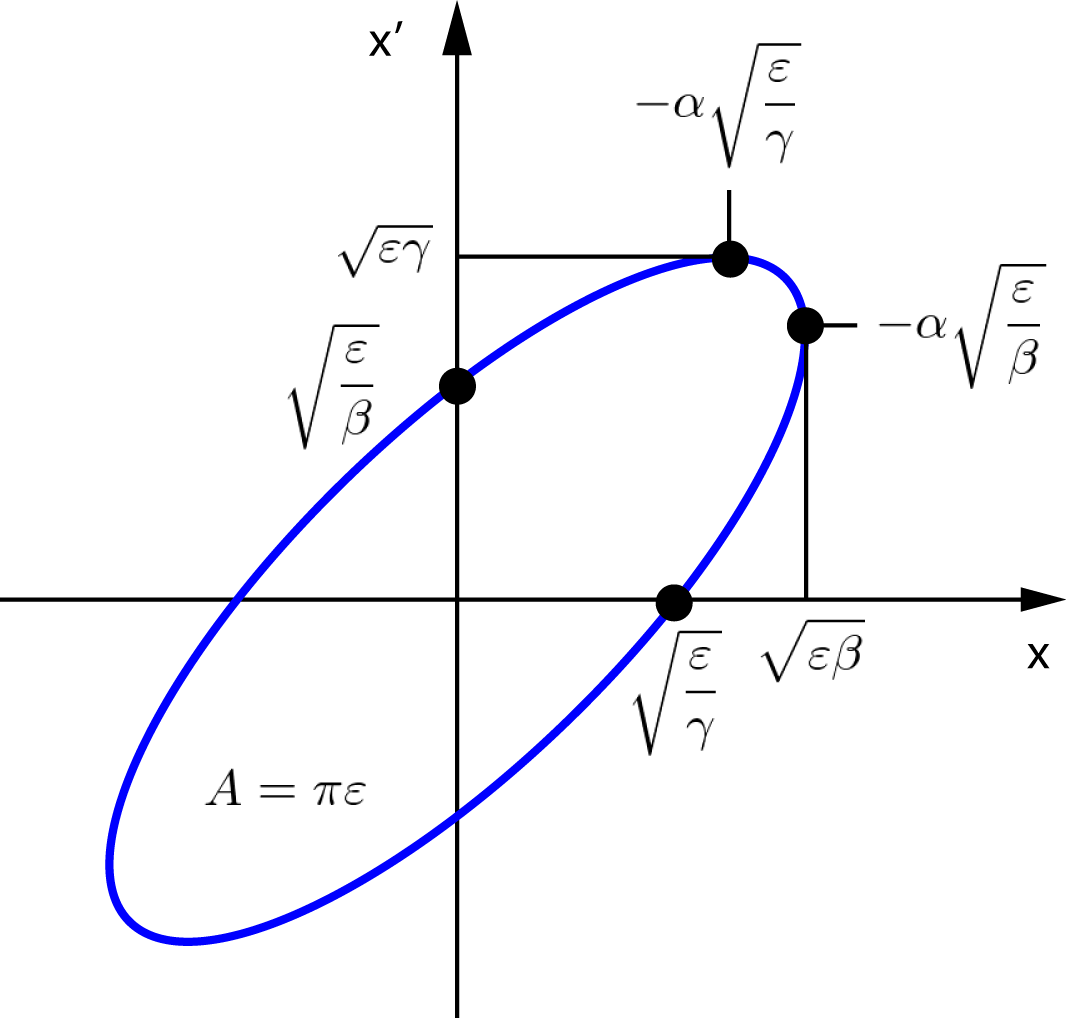}
\caption{Rms-emittance ellipse defined using the beam matrix (left) and the generic Courant-Snyder emittance ellipse (right).}
\label{fig:rmsemittanceellipse}
\end{figure}

For simplicity let us assume phase space distributions centred in the origin of the coordinate system
\begin{align}
x_i &= X_i-\langle X \rangle \label{eq:xnorm},\\
{x'}_i &= X'_i-\langle X' \rangle \label{eq:xpnorm}.
\end{align}
The three terms that we need for the definition of the rms-emittance are:
\begin{align}
{x_{rms}}^2 &= \langle x^2 \rangle = \frac{1}{N}\sum_{i=1}^{N} {x_i}^2 \label{eq:x2rms},\\
{{x'}_{rms}}^2 &= \langle {x'}^2 \rangle = \frac{1}{N}\sum_{i=1}^{N} {x'}_i^2 \label{eq:xp2rms},\\
{xx'}_{rms} &= \langle {xx'} \rangle = \frac{1}{N}\sum_{i=1}^{N} {x}_i{x'}_i \label{eq:xxprms}.
\end{align}
We then use these terms to define the \emph{beam matrix}
\begin{align}
\Sigma=
\begin{bmatrix}
\Sigma_{11} & \Sigma_{12} \\
\Sigma_{12} & \Sigma_{22}
\end{bmatrix}=
\begin{bmatrix}
\langle {x}^2 \rangle & \langle {xx'} \rangle \\
\langle {xx'} \rangle & \langle {x'}^2 \rangle \\
\end{bmatrix} \label{eq:beammatrix},
\end{align}
the rms-emittance is defined as the determinant of the beam matrix.
\begin{align}
\varepsilon_{rms}= \sqrt{\Sigma_{11} \Sigma_{22} -
\Sigma_{12}^2}= \sqrt{\det \Sigma} \label{eq:rmsemittance}.
\end{align}
The beam emittance is usually associated to the area of the smallest ellipse in phase-space containing the particles. In case of the rms-emittance the associated ellipse is defined by the beam matrix as shown in \Fref{fig:rmsemittanceellipse} where
\begin{align}
\tan 2\varphi= \frac{2\Sigma_{12}}{\Sigma_{22}-\Sigma_{11}} \label{eq:rmemittancephi}.
\end{align}

Courant and Snyder developed a formalism for description of the particle dynamics in the transverse space~\cite{Courant:1997rq} where the emittance ellipse is described by four parameters, also known as the Twiss parameters: $\varepsilon, \alpha, \beta$ and $\gamma$ shown in \Fref{fig:rmsemittanceellipse}
where $\varepsilon$ is the emittance and
\begin{align}
\gamma= \frac{1+\alpha^2}{\beta} \label{eq:csgamma}.
\end{align}

The Courant-Snyder parameters are closely related to the beam matrix~\cite[p.~161]{Wiedemann:2015fja}
\begin{align}
\Sigma=
\begin{bmatrix}
\Sigma_{11} & \Sigma_{12} \\
\Sigma_{12} & \Sigma_{22}
\end{bmatrix}=
\varepsilon
\begin{bmatrix}
\beta & -\alpha \\
-\alpha & \gamma \\
\end{bmatrix} \label{eq:csbeammatrix}.
\end{align}
The equation of the emittance ellipse with the Courant-Snyder formalism is
\begin{align}
    \varepsilon= \gamma\,x^2+2\,\alpha\,x\,x'+\beta\,{x'}^2 \label{eq:emittanceelipse},
\end{align}
and the area is simply $A=\pi\,\epsilon$. Note that the projected width of the ellipse along $x$, the beam size, depends only on the emittance and the $\beta$ parameter
\begin{align}
    \sigma= \sqrt{\varepsilon\,\beta} \label{eq:betaepsilon}.
\end{align}
As explained in previous lectures in this school the $\beta$ parameter is the value of the betatron function at our location. 

\section{Adiabatic damping}
In the introduction it was explained how the emittance is an invariant parameter of the particles motion in phase-space. In reality this invariance is only valid under well defined circumstances, in particular it is not valid if the particles are accelerated (change of longitudinal momentum).
The $x'$ coordinate of the phase-space is the angle the trajectory of the particle forms with the longitudinal axis. This angle can be assumed as the ratio between the transverse and the longitudinal momenta.

If the beam is accelerated or decelerated the $x'$ axis of the phase-space is contracted or expanded respectively, leading to a change in emittance.
If a beam is accelerated from momentum $p_1$ to momentum $p_2$ the new emittance will thus be
\begin{align}
    \varepsilon_2= \varepsilon_1 \frac{p_1}{p_2} \label{emittanceacceleration},
\end{align}
where the relativistic momentum $p$ is
\begin{align}
    p= m_0\,\beta_\text{rel}\, \gamma_\text{rel}\, c \label{relmomentum}.
\end{align}
For this reason the values of the beam emittances are usually normalized by the relativistic factor, this new quantity, usually indicated with $\varepsilon^*$ or $\varepsilon_\text{N}$, is called \emph{normalized} emittance
\begin{align}
    \varepsilon^*= \varepsilon_\text{geo}\, \beta_\text{rel}\, \gamma_\text{rel},
\end{align}
where $\varepsilon_\text{geo}$, the geometrical emittance, is just the emittance that we have defined before and that we can physically measure.

The advantage of the normalized emittance is that it is invariant also respect to the acceleration~\cite[p.~60]{Lee:2019xas} and it is used as a figure of merit of the quality of a beam along the whole acceleration cycle, often spanning over different accelerators and related transfer lines.

\section{Effects of dispersion}
Particles of different momentum experience different forces while traversing magnetic fields. For this reason particles with different momentum will describe different trajectories along the accelerator or transfer line.
The dispersion is a property of the magnetic structure and its value depends on the position, for this reason we usually talk of a dispersion function~\cite[p.~236]{Wiedemann:2015fja}.

\begin{figure}[htb]
\centering
\includegraphics[align=c, width=0.4\linewidth]{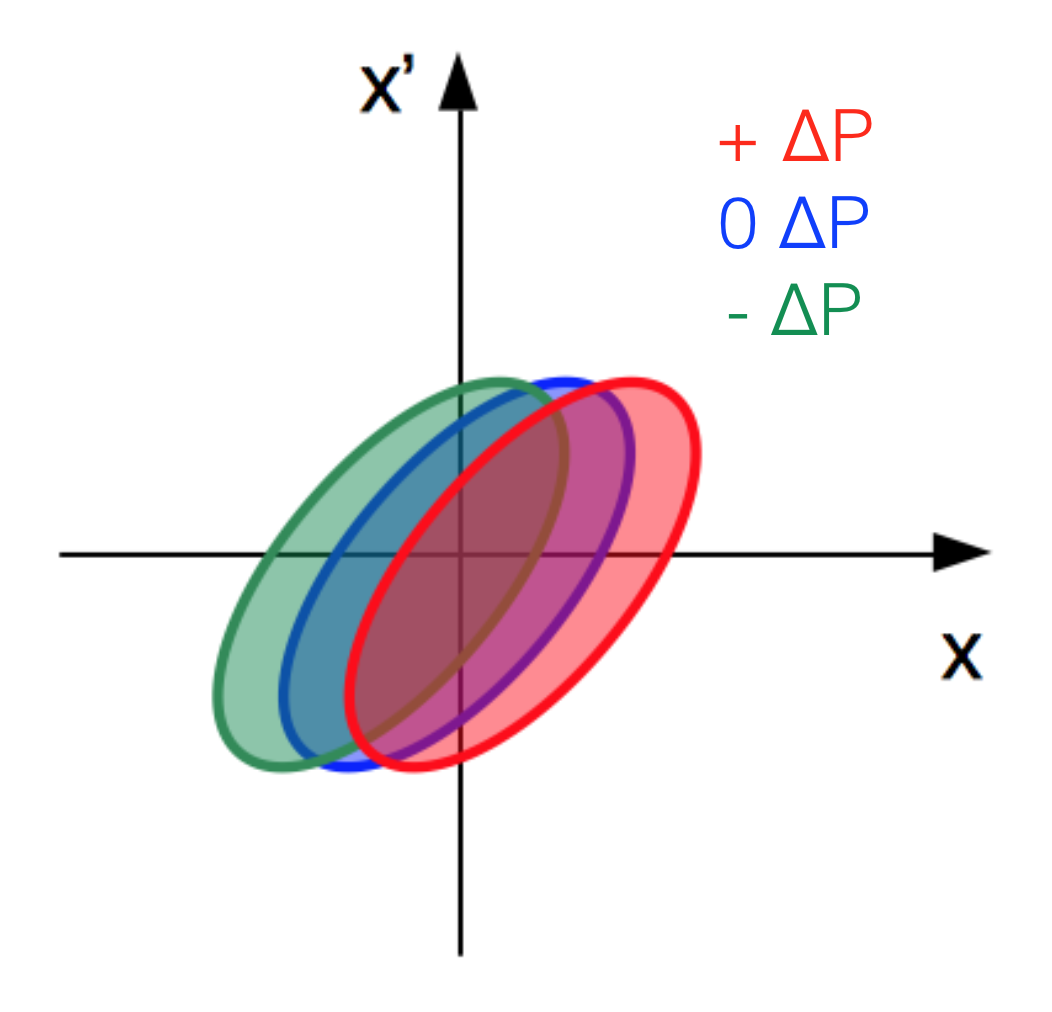}
\caption{Particles with a momentum error describe ellipses in phase space with an horizontal offset proportional to the momentum error and the dispersion value.}
\label{fig:dispersion}
\end{figure}

The value of the dispersion at one location is defined as the offset in trajectory of a particle with a momentum error w.r.t. the same particle without momentum error, normalized by the momentum error itself
\begin{align}
   D=\frac{\Delta x}{\frac{\Delta p}{p}} \label{eq:dispersion}.
\end{align}
Looking at the phase-space distribution at a location with finite dispersion we see that off-momentum particles belong to emittance ellipses shifted horizontally from the on-momentum ellipse, as sketched in \Fref{fig:dispersion}. Clearly the area, and thus the emittance, of this distribution is larger than the area of one single ellipse, corresponding to the distribution in case of zero dispersion.

Since the dispersion varies location by location we could conclude that the emittance is no longer invariant. In reality the conservation of the phase-space should include all the coordinates and we should talk about the conservation of the 6D hyper-volume of phase-space $(x,x',y,y',t,p)$. In this case we can just consider a 3D extension of our phase-space, adding the particle momentum as third axis, the particles distribution in phase-space is now an ellipsoid that is deformed by the dispersion, but the volume is conserved~\cite[p.~60]{Lee:2019xas}.

In a simplified vision the volume can be decomposed into thick, overlapping, ellipses each one referring to a discrete momentum error. As the dispersion changes these ellipses shift horizontally, but the sum of the volumes remains the same.

Clearly measuring the emittance in presence of dispersion and momentum error is very complicated because of the need to determine the position, the angle and the momentum of the particles at the same time. In a real beam a finite momentum spread is unavoidable and the best solution is to measure the emittance at locations where the dispersion is zero. If the dispersion and the momentum spread are well known, it is in principle possible to deconvolute the momentum spread contribution from the sampled phase-space distributions.

\section{Importance of the transverse emittance}
In many cases we want to know the size of the beam in our accelerator, first of all to know if the beam has sufficient clearance inside the vacuum chamber, i.e. if the machine has sufficient aperture, but also to exploit a facility efficiently. The beam sizes can be expressed in terms of the emittance, a property of the beam, and the betatron function, a property of the machine
\begin{align}
    \sigma=\sqrt{\beta \varepsilon}.
\end{align}

In case of colliders the aim is to maximize the luminosity, that is the rate at which collisions take place. This is inversely proportional to the beam size at the collision point
\begin{align}
    L= \frac{N_{b1}\,N_{b2}\,f_\text{rev}\,k_b}{2\,\pi\,\sqrt{(\sigma^2_{x1}+\sigma^2_{x2})(\sigma^2_{y1}+\sigma^2_{y2})}} \label{eq:luminosity},
\end{align}
where $N_{b\,i}$ are the bunch populations of the two colliding beams, $f_\text{rev}$ the revolution frequency, $k_b$ the number of colliding bunches and $\sigma$ the beam sizes for the two beams and two planes
\begin{align}
    \sigma_i= \sqrt{\varepsilon_i\,\beta_i} \quad i \in [x_1, y_1, x_2, y_2] \label{eq:beamsizes}.
\end{align}
If the two beams are equal the luminosity is proportional to $1/(\sigma_x\,\sigma_y)$ or $1/\sqrt{\varepsilon_x\,\varepsilon_y}$.

In synchrotron light sources the spectral brightness of the emitted radiation depends on the beam brightness
\begin{align}
    \bar{B}= \frac{2I}{\pi^2\varepsilon_x\,\varepsilon_y} \label{eq:brightness},
\end{align}
where $I$ is the beam current and $\varepsilon_{x,y}$ are the transverse emittances. In modern facilities, where a high spatial coherence of the radiation is requested, the control and minimization of the emittance are the main challenges.

\section{Phase space evolution}
\begin{figure}[htb]
\centering
\includegraphics[align=c, width=0.8\linewidth]{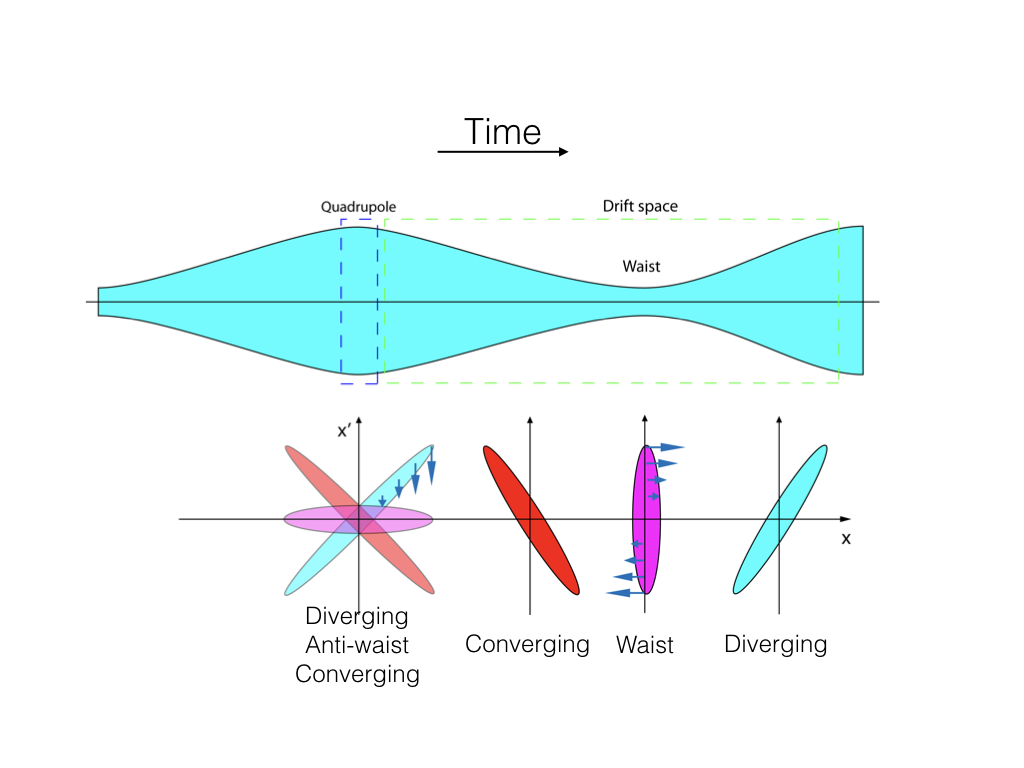}
\caption{Evolution of the phase space distribution in drift spaces and quadrupoles. In drift spaces the ellipses are sheared horizontally, while inside quadrupoles are sheared vertically as indicated by the arrows.}
\label{fig:phasespaceevolution}
\end{figure}

\Figure[b]~\ref{fig:phasespaceevolution} illustrates the evolution of the phase space ellipses in drift spaces and inside quadrupoles. In drift spaces the ellipses are sheared horizontally, while inside quadrupoles are sheared vertically. If the ellipse is tilted backward (red ellipse in the figure) the shear transformation will reduce the tilt as the position (time) advances until it will eventually vanish and the ellipse will be upright. This describes a converging beam that ends in a waist (local minimum in transverse size). After the waist the ellipse continues to deform in the same way, but this time the tilt will be forward (cyan ellipse in the figure) and could increase indefinitely with the major axis becoming longer and longer, the beam is now diverging. When the beam enters a quadrupole the magnetic forces will kick the particles proportionally to their offset from the centre, this results in a vertical shear. If the quadrupole is focusing the right side of the phase-space will move downward and the left side upward, the effect will be the opposite for a defocusing quadrupole. The tilt of a diverging beam inside of a focusing quadrupole will reduce until the ellipse is horizontal (anti-waist, local maximum in beam size), will then become negative and increase until the end of the quadrupole~\cite[p.~221]{Wiedemann:2015fja}.

\begin{figure}[htb]
\centering
\includegraphics[align=c, width=0.6\linewidth]{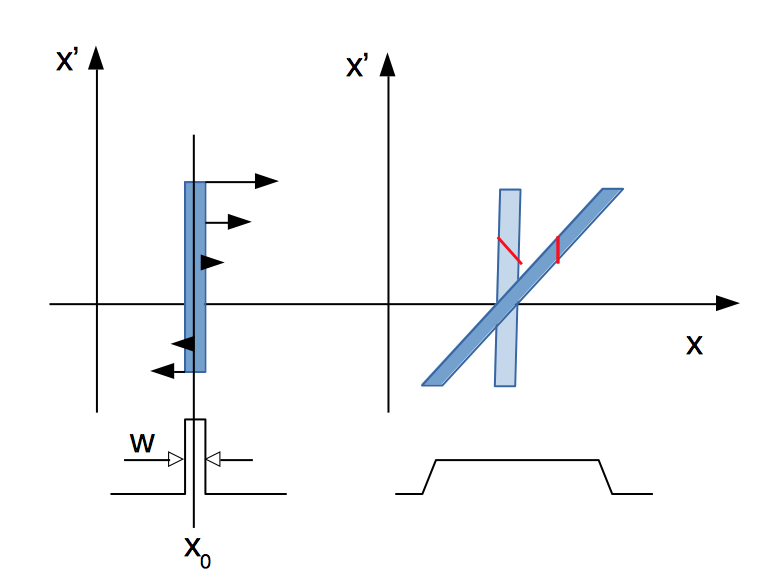}
\caption{Evolution in phase-space of a thin vertical rectangle in a drift space. The bottom graphs indicate the projections along $x$ i.e. the profiles.}
\label{fig:slitphasespacedrift}
\end{figure}

A very interesting case, for our purpose, is the evolution in phase-space of a thin vertical rectangle in a drift space as shown in \Fref{fig:slitphasespacedrift}. The points of the phase-space evolve according to this simple linear transformation 
\begin{align}
    x_2 &=x_1+x_1'\,L \label{eq:xdrift},\\
    x_2' &= x_1' \label{eq:xpdrift},
\end{align}
where $(x_1, x_1')$ and $(x_2, x_2')$ are the coordinate of the particle before and after the drift space and $L$ is the length of the drift space.

The profiles of the distributions are
\begin{align}
    p_1(x) &= \int i_1(x, x')\,dx' \label{eq:slitprofilebeforedrift},\\
    p_2(x) &= \int i_2(x, x')\,dx' = \int i_1(x-x'\,L, x')\,dx'\label{eq:slitprofileafterdrift}.
\end{align}
If the width of the rectangle $w$ is reduced to zero, using equation \eqref{eq:slitprofileafterdrift}, we can write
\begin{align}
    \lim_{w \to 0}
    \begin{cases}
    i_1(x, x')&= \delta(x-x_0)\xi(x')\\
    p_2(x)&= i_1(x_0, \frac{x-x_0}{L})= \xi(\frac{x-x_0}{L})
    \end{cases} \label{eq:slitdrifttozero},
\end{align}
this means that if the profile $p_2(x)$ is known then the angular distribution $\xi(x')$, and thus $i_1(x_0, x')$, can be calculated.

\subsection{Transport matrices}
In the previous section we have seen how, in a linear system made of quadrupoles and drift spaces, points of phase-space can be mapped from one location to another. Equations \eqref{eq:xdrift} and \eqref{eq:xpdrift} can be rewritten in matrix notation and extended to multiple segments
\begin{align}
    \begin{bmatrix}
    x_1\\x_1'
    \end{bmatrix}
    &= M_1
    \begin{bmatrix}
    x_0\\
    x_0'
    \end{bmatrix} ,\\
    \begin{bmatrix}
    x_2\\x_2'
    \end{bmatrix}
    &= M_2
    \begin{bmatrix}
    x_1\\
    x_1'
    \end{bmatrix},\\
    \begin{bmatrix}
    x_3\\x_3'
    \end{bmatrix}
    &= M_3
    \begin{bmatrix}
    x_2\\
    x_2'
    \end{bmatrix},
\end{align}
and with simple algebra
\begin{align}
    \begin{bmatrix}
    x_3\\x_3'
    \end{bmatrix}
    = M_3\,M_2\,M_1
    \begin{bmatrix}
    x_0\\
    x_0'
    \end{bmatrix}
    = M_{0\Rightarrow 3}
    \begin{bmatrix}
    x_0\\
    x_0'
    \end{bmatrix} \label{eq:matrixcombination}.
\end{align}

The $M$ matrices are the transport matrices and can be easily derived for the linear elements~\cite[p.~44]{Lee:2019xas}
\begin{align}
    M_\text{Drift}=     
    \begin{bmatrix}
    1 & L\\
    0 & 1
    \end{bmatrix}
    \quad
    M_\text{Quad short}=     
    \begin{bmatrix}
    1 & 0\\
    -\frac{1}{f} & 1
    \end{bmatrix} \label{eq:transportmatrices},
\end{align}
where $L$ is the length of the drift space and $f$ the focal length of the thin, or short, quadrupole. For real, long, quadrupoles the matrices are a bit more complicated
\begin{align}
    M_\text{QF} &=     
    \begin{bmatrix}
    \cos(\sqrt{k}\,L_Q) & \frac{1}{\sqrt{k}}\sin(\sqrt{k}\,L_Q)\\
    -\sqrt{k}\sin(\sqrt{k}\,L_Q) & \cos(\sqrt{k}\,L_Q)
    \end{bmatrix} \label{eq:qftransportmatrix},
\end{align}
\begin{align}
    M_\text{QD} &=     
    \begin{bmatrix}
    \cosh(\sqrt{-k}\,L_Q) & \frac{1}{\sqrt{-k}}\sinh(\sqrt{-k}\,L_Q)\\
    \sqrt{-k}\sinh(\sqrt{-k}\,L_Q) & \cosh(\sqrt{-k}\,L_Q)
    \end{bmatrix} \label{eq:qdtransportmatrix},
\end{align}
where $L_Q$ is the magnets length and $k$ is the effective focusing strength of the quadrupole
\begin{align}
    k= \frac{1}{B\rho} \frac{\partial B_y}{\partial x}
    \quad \text{or (the same)} \quad
    k= \frac{1}{B\rho} \frac{\partial B_x}{\partial y}
    \label{eq:quadk},
\end{align}
with $B\rho$ the rigidity of the beam which is just proportional to the momentum. The value of $k$ is positive for a focusing quadrupole and negative for a defocusing quadrupole. In fact a quadrupole is always focusing in one plane and defocusing in the other.
For a short quadrupole the matrix in \eqref{eq:transportmatrices} can be used with $f=k L_Q$, where $f$ is negative for a defocusing quadrupole.

The transport matrix of a complex beam line, even of a whole ring, can be calculated by multiplying the matrices of each linear element as shown in \eqref{eq:matrixcombination}.
If we can transport each point of the phase-space it must be possible to transport also the ellipses that we associate with the emittance and thus obtain the Twiss parameters $\alpha$, $\beta$ and $\gamma$ at any new location.

Assuming the following single particle transport matrix
\begin{align}
    \begin{bmatrix}
    x_1\\x_1'
    \end{bmatrix}
    =
    \begin{bmatrix}
    c & s\\
    c' & s'
    \end{bmatrix}
    \begin{bmatrix}
    x_0\\
    x_0'
    \end{bmatrix} \label{eq:cstransportmatrix},
\end{align}
with some calculation we obtain the following result for the Twiss parameters
\begin{align}
    \begin{bmatrix}
    \beta_1\\
    \alpha_1\\
    \gamma_1
    \end{bmatrix}
    =
    \begin{bmatrix}
    c^2 & -2\,c\,s & s^2\\
    -c\,c' & c\, s' + c'\,s & -s\, s'\\
    {c'}^2 & -2\,c'\,s' & {s'}^2
    \end{bmatrix}
    \begin{bmatrix}
    \beta_0\\
    \alpha_0\\
    \gamma_0
    \end{bmatrix} \label{eq:twisstransportmatrix}.
\end{align}

\subsection{Phase space mismatch} \label{phasespacemismatch}
In circular accelerators, or storage rings, the stability conditions and the closure of the orbit require that the Twiss parameters transported over one turn, with equation \eqref{eq:twisstransportmatrix}, remain the same. In order to fulfil this conditions the matrix must have an eigenvalue equal to one and the Twiss vector is then just the corresponding eigenvector. In other words the Twiss parameters, defining the evolution of the single particle around the ring, derive just from the design of the magnetic lattice.

Until now we have referred to the phase-space distribution as a property of the beam, which is correct, but now we see that the lattice itself defines the parameters of the ellipse bar the emittance. How do these two aspects combine? If we inject a beam that has a phase space distribution (beam envelope) whose ellipse is different from the one defined by the Twiss parameters of the lattice, the beam envelope will rotate turn after turn inscribing the ellipse defined by the lattice. The amplitude (emittance) of this inscribed ellipse is the one that contains the injected beam envelope~\cite[v.~1,~p.~240]{Turner:1994bd}.

\begin{figure}[htb]
\centering
\includegraphics[align=c, width=0.4\linewidth]{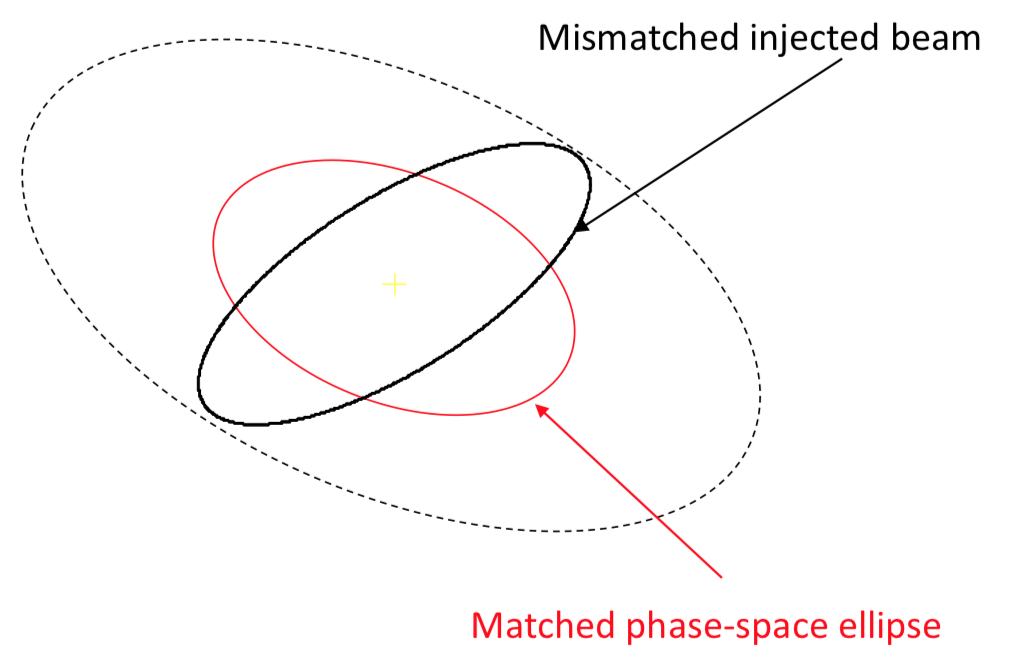}
\caption{Mismatched injection. The dotted ellipse represents the lattice ellipse that contains the mismatched injected beam, the red ellipse represents a matched beam while the black ellipse represents a mismatched beam. The black ellipse will rotate, turn after turn, inscribing the dotted ellipse. Due to the non linearities of the magnetic structure the particles will eventually diffuse and fill the entire dotted ellipse.}
\label{fig:mismatch}
\end{figure}

Due to non linearities the particles evolve around the ellipse at different speeds so that after a long time the initial beam ellipse will diffuse and cover entirely the inscribed ellipse. As a consequence the final emittance will be larger than the injected one. In applications where the preservation of the emittance is important, like in hadron accelerators, the matching between rings and transfer lines is very important. \Figure[b]~\ref{fig:mismatch} shows a sketch of the injection mismatch mechanism.
If a turn by turn profile monitor is available it is possible to identify a mismatched injection by the oscillation of the measured beam size turn after turn. These oscillation will have a frequency that is twice that of the tune of the machine.

\section{Measurement of the transverse emittance}
Until now we have seen what the emittance is and why it is important for the exploitation of particle accelerators. In order to measure the emittance we have two possibilities:
\begin{itemize}
    \item Direct sampling of the phase space
    \item Sampling of the real space, often just the projections along $x$ or $y$, and use the beam dynamic theory to infer the emittance
\end{itemize}
The first method is the preferred one for low energy beams, after the source and in the first stages of acceleration, where the phase-space distribution is far from the bivariate normal distribution due to space charge. The second possibility groups a number of techniques:
\begin{itemize}
    \item Single profile measurement, requires the knowledge of the Courant-Snyder (Twiss) parameters, often used in circular accelerators
    \item Multiple profile measurement, requires knowing the transport matrices between the profile monitors and solves the emittance and the Twiss parameters, often used in transfer lines
    \item Quadrupolar scan, requires knowing the transfer function of the quadrupole, like for the previous method solves both the emittance and the Twiss parameters, mainly used in linacs
\end{itemize}

When talking about emittance measurement one aspect becomes evident, different people use different definitions for the emittance. It is thus important to define what we mean with the different nomenclature. We have already defined the rms-emittance in terms of the moments of the distributions and we know that the emittance is linked to the area of an ellipse
\begin{align}
    \varepsilon_\text{rms}= \frac{A}{\pi} ,
\end{align}
similarly we can define the two-rms-emittance as the area of an ellipse scaled by a factor two in both dimensions divided by $\pi$ so that
\begin{align}
     \varepsilon_\text{2rms}= 4  \varepsilon_\text{rms}.
\end{align}
Similarly we can define the $90$\% emittance as 
\begin{align}
    \varepsilon_\text{90\%} = \frac{A_\text{90\%}}{\pi}
\end{align}
were $A_\text{90\%}$ is the area of the Twiss ellipse that contains 90\% of the particles. Similarly 95\% emittance and 85\% emittance are defined.

In case of Gaussian distributions it is relatively easy to convert from one to the other, but in case the distributions are not Gaussian it is not so easy. The reason to have all these definitions is that when the distributions are not Gaussian it is important to define the quantity that better describes the quality of the beam which can differ case by case (do we care about the tails? is the core the important part? etc.).

\subsection{Slit and grid method}
We have seen how, in a particular case, the knowledge of a transverse beam profile can be used to infer the angular distribution of particles in phase space.
This principle is used when direct phase-space sampling is required, for example when the particle distribution is dominated by space-charge and is not Gaussian.

There are several techniques that exploit this principle, the simplest is the so called \emph{slit and grid} method. In this case a solid blade, sufficiently thick to stop the beam, is placed on the path of the particles. On the blade a thin slit is cut either horizontally or vertically so that only the particles with a well defined position can pass trough. At an adequate distance downstream of the blade a profile monitor is installed, like a scintillating screen or a wire harp (the grid part of the name comes from \emph{wire grid}). The blade is then moved in small steps perpendicularly to the slit while the profiles of the emerging beamlets are recorded at every step. Each profile, after the opportune scaling described by equation \eqref{eq:slitdrifttozero}, corresponds to the angular distribution of the particles for a given position, when all the data is combined this corresponds to the sampling of the whole phase-space~\cite[p.~702]{Chao:2013rba}.

\begin{figure}[htb]
\centering
\includegraphics[align=c, width=0.4\linewidth]{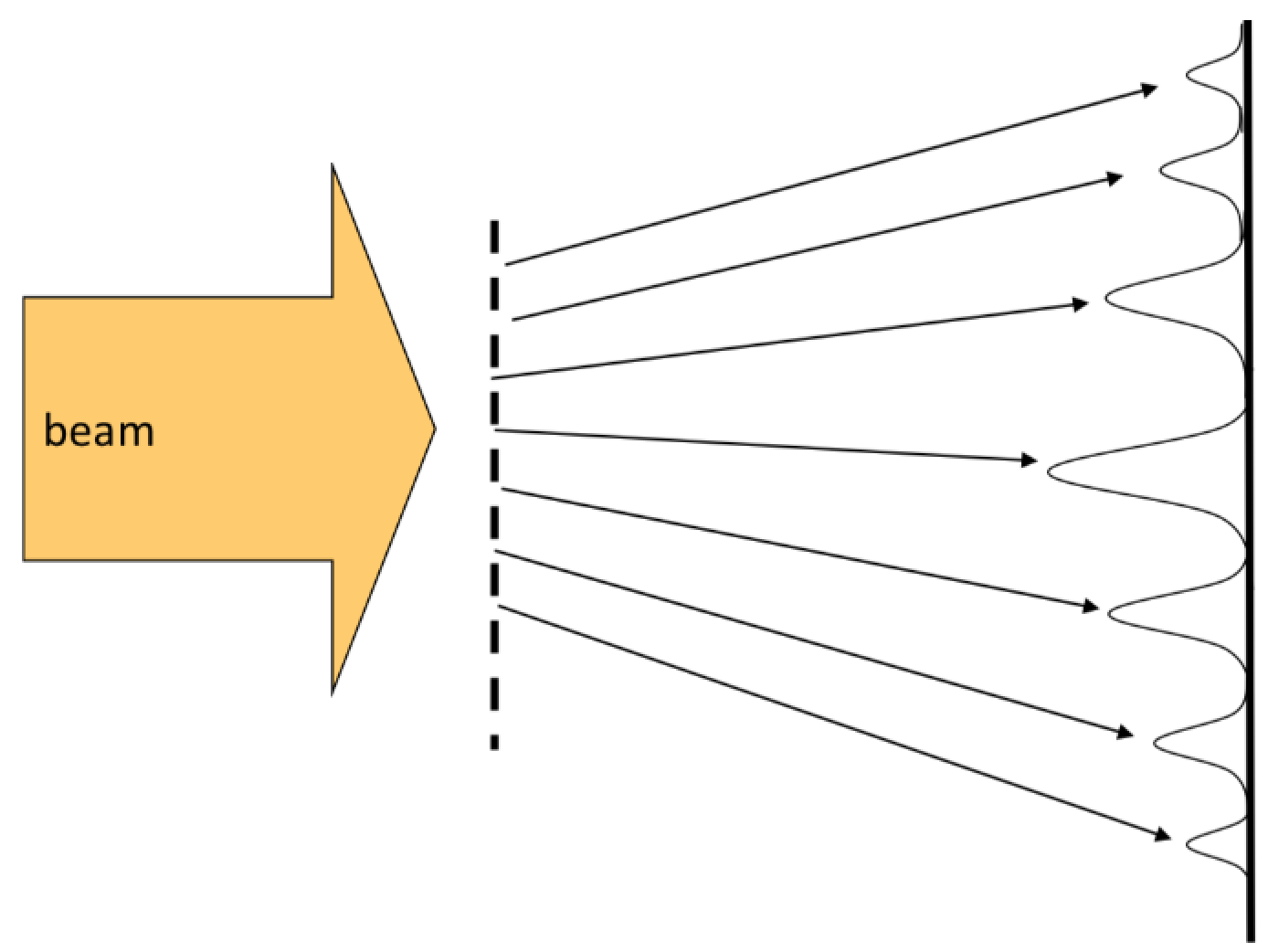}
\caption{Sketch of a multi slit emittance meter. The beam is stopped by the solid blade and only thin beamlets, in correspondence of the slits, are transmitted. The profiles of these beamlets, acquired downstream of the blade, allow the reconstruction of the phase-space distribution at the location of the blade.}
\label{fig:multislit}
\end{figure}

Instead of a single slit it is possible to cut several parallel slits so that the sampling can be performed in parallel instead of scanning the blade. In this case a 2D profile monitor is required in order to separate the profiles of the beamlets coming from the different slits. It is also crucial that these profiles do not overlap so a minimum distance between slits has to be ensured. Often these multi-slit systems are also scanned to increase the resolution of the sampling, but result in a faster measurement than a single slit. \Figure[b]~\ref{fig:multislit} illustrates the slit and grid method of a multi-slit system.

In a real system the slit will always have a finite dimension, so that the measured profiles will be the result of the convolution of the slit width and the angular distribution. The effect of the slit width is influenced by the original beam divergence and the distance between the profile monitor and the blade. The resolution of the profile monitor is also important and has to be sufficient for measuring accurately the profiles of the beamlets. A thin slit means a weak signal in the profile monitor and a distortion due to a large fraction of particles scattered by the edge of the slit. On the contrary a wide slit means a large error on the calculation of the angular distribution. All these parameters have to be carefully considered and optimized on a case by case basis. The slit and grid method is usually limited to low energy beams since the blade must be able to stop completely the particles. In case of high energy the thickness of the blade would filter particles not only on the position, but also on the angle, moreover scattering on the slit edges would introduce errors in the angular distribution.

\subsection{The pepper pot}
The multi-slit method described above can be extended further by replacing the slits with small holes in a rectangular pattern and using a high resolution scintillating screen. In this case it is possible to compute the horizontal and vertical profiles of the beamlets from the same 2D image. With a pepper pot device it is thus possible to sample the entire phase-space in a single shot~\cite{Kremers:2013cv}.
Although very attractive, this technique is quite difficult to implement and requires a complex analysis of the images. The phase-space sampling resolution is limited by the need to keep the beamlets well separated. The spot corresponding to each hole has to be isolated and then the profiles corresponding to every row or column have to be reconstructed. 

\subsection{Emittance from multiple profiles}
In case we can describe the distribution in phase space with the Courant-Snyder parameters (ellipses), we can easily calculate the transport matrices of the distribution along a drift space. In particular for any location we can use equation \eqref{eq:twisstransportmatrix} and write
\begin{align}
    \beta_1= 
    \begin{bmatrix}
    c_1^2 & -2\,c_1\,s_1 & s_1^2
    \end{bmatrix}
    \begin{bmatrix}
    \beta_0\\
    \alpha_0\\
    \gamma_0
    \end{bmatrix} \label{eq:batatransport}.
\end{align}
If we write this equation for three locations and multiply left and right by the emittance $\varepsilon$ we obtain
\begin{align}
    \begin{bmatrix}
    \varepsilon \beta_1\\
    \varepsilon \beta_2\\
    \varepsilon \beta_3
    \end{bmatrix}
    = \varepsilon
    \begin{bmatrix}
    c_1^2 & -2\,c_1\,s_1 & s_1^2 \\
    c_2^2 & -2\,c_2\,s_2 & s_2^2 \\
    c_3^2 & -2\,c_3\,s_3 & s_3^2
    \end{bmatrix}
    \begin{bmatrix}
    \beta_0\\
    \alpha_0\\
    \gamma_0
    \end{bmatrix}
    = \varepsilon
    M
    \begin{bmatrix}
    \beta_0\\
    \alpha_0\\
    \gamma_0
    \end{bmatrix}
    \label{eq:threescreensmatrix}.
\end{align}
Using equation \eqref{eq:betaepsilon} and using three profile monitors to measure the beam sizes at the chosen locations $1$, $2$ and $3$ we can solve the system of equations and calculate the Twiss parameters and the emittance at location $0$~\cite[v.~1,~p.~245]{Turner:1994bd}
\begin{align}
    M^{-1}
    \begin{bmatrix}
    \sigma_1^2\\
    \sigma_2^2\\
    \sigma_3^2
    \end{bmatrix}
    =
    \begin{bmatrix}
    a\\b\\c
    \end{bmatrix}
    = \varepsilon
    \begin{bmatrix}
    \beta_0\\
    \alpha_0\\
    \gamma_0
    \end{bmatrix} ,
\end{align}
\begin{align}
    \begin{cases}
    a &= \varepsilon \,\beta_0\\
    b &= \varepsilon \,\alpha_0\\
    c &= \varepsilon \,\gamma_0\\
    \gamma_0 &= \frac{1+\alpha_0^2}{\beta_0}
    \end{cases}
    \quad \Longrightarrow \quad
    \begin{cases}
    \varepsilon &= \sqrt{a\,c-b^2} \\
    \beta_0 &= \frac{a}{\varepsilon} \\
    \alpha_0 &= \frac{b}{\varepsilon} \\
    \gamma_0 &= \frac{c}{\varepsilon}
    \end{cases} \label{eq:threescreenssolution}.
\end{align}
In order to solve the system of equations the matrix $M$ of equation \eqref{eq:threescreenssolution}, that depends on the individual transport matrices, must be invertible. The choice of the locations 1, 2 and 3 has also an impact on the sensitivity on the measurement errors. Usually the three profile monitors are located one in a converging beam section before the waist, one in or near the waist and one after the waist in the diverging section. The layout of such a measurement line, in particular the distance between the profile monitors, has to be optimized on a case by case basis.

\subsection{Quadrupole strength scan}
We have just seen how the phase-space distribution can be inferred from multiple beam size measurements using the transport matrices corresponding to the segments between the profile monitors. In fact we can extend this concept and instead of using multiple profile monitors use a single one, but change the transport matrix from a reference location upstream and the monitor itself~\cite[p.~226]{Wiedemann:2015fja}.

In practice this is accomplished by installing the profile monitor downstream of a focusing quadrupole and changing the strength of the magnetic field gradient in the magnet ($k$). Clearly this method can only be used in transfer lines or linacs and is not applicable to circular machines. By changing the focusing field in three precise steps we can write three transport equations and solve the system like for the three screens method, usually the reference point is taken just before the quadrupole.

\begin{figure}[htb]
\centering
\includegraphics[align=c, width=0.7\linewidth]{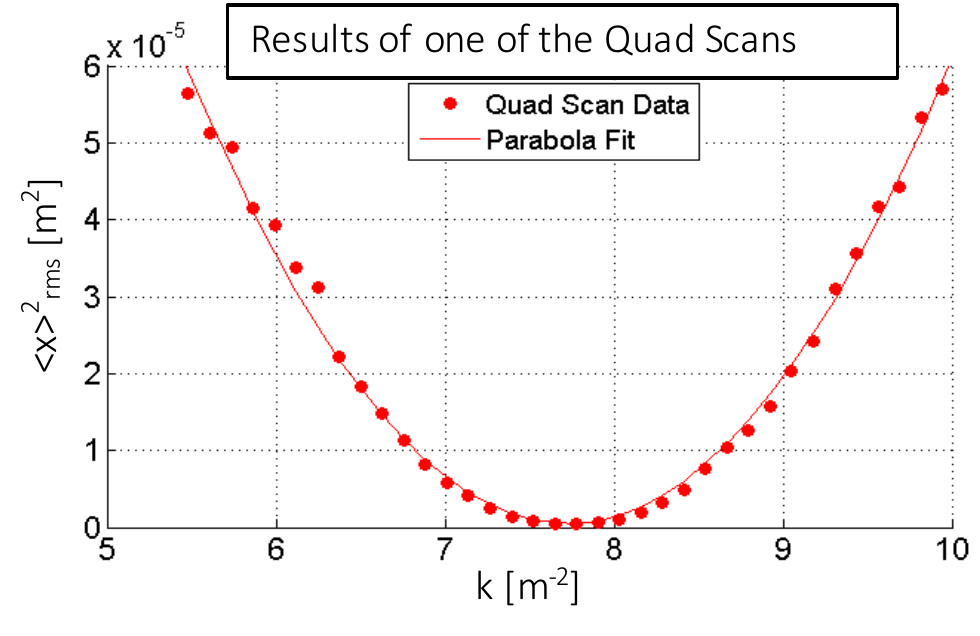}
\caption{Result of a quadrupole strength scan or \emph{quadscan}. The measured beam widths squared are plotted against the magnet strength, then a parabola is fitted. From the coefficients of the parabola and the geometry of the beam line it is possible to derive the beam emittance and the Twiss parameters.}
\label{fig:quadscanparabola}
\end{figure}

One big advantage of this method is that we are not limited to three focusing values, we can in fact make as many steps we want with just a little extra time needed. Also in the case of the multiple screens we could use more than three, but that would require the addition of expensive hardware.
With more that three measurements however the problem is over constrained, we have more equations than unknowns. In this case instead of solving a system we perform a minimisation.
The square of the measured beam size at the monitor location plotted against the quadrupole strength will describe a parabola with the minimum for the strength at which the waist of the beam is located at the profile monitor. In reality measurement errors and other perturbations will disperse the points around the parabola.
For each value of $k$ we can calculate the transport matrix of equation \eqref{eq:cstransportmatrix} and apply equation \eqref{eq:batatransport}
\begin{align}
    \varepsilon\beta_i= \sigma_i^2= c_i^2\beta_0\,\varepsilon - 2\,c_i s_i \alpha_0\,\varepsilon + s_i^2 \gamma_0\,\varepsilon \label{eq:quadsacanbase},
\end{align}
where $\sigma_i$ are the measured beam sizes and the coefficients $c_i$ and $s_i$ depend on $k$.
For a short quadrupole of strength $k$ and length $\ell$, a distance between the quadrupole and the profile monitor of $d$ and using the matrices in \eqref{eq:transportmatrices}, equation \eqref{eq:quadsacanbase} can be written as
\begin{align}
    \sigma^2(k)= a\,k^2 - 2\,a\,b\,k+a\,b^2+c \label{eq:quadscank}.
\end{align}
The coefficients $a$, $b$ and $c$ are obtained by minimising the difference between the measured beam sizes $\sigma_i$ and the predicted one $\sigma(k_i)$ (residuals). In other words by fitting the function in \eqref{eq:quadscank} to the experimental data points $(\sigma_i, k_i)$ as shown in \Fref{fig:quadscanparabola}. 
With some manipulations it is possible to solve for the emittance and the Twiss parameters
\begin{align}
    \varepsilon &= \frac{\sqrt{a\,c}}{d^2\ell^2} ,\\
    \varepsilon\beta_0 &= \frac{a}{d^2\ell^2} ,\\
    \varepsilon\alpha_0 &= -\frac{a}{d^2\ell^2}(b\,\ell-\frac{1}{d}) ,\\
    \varepsilon\gamma_0 &= \frac{c}{d^2}+\frac{a}{d^2\ell^2}(b\,\ell-\frac{1}{d}).
\end{align}

\subsection{Emittance measurement in circular machines}
In section \ref{phasespacemismatch} we have seen how, in a circular machine, the transport of the Twiss parameters over one full turn must yield the same Twiss vector
\begin{align}
    \begin{bmatrix}
    \beta(z)\\
    \alpha(z)\\
    \gamma(z)
    \end{bmatrix}
    =
    \begin{bmatrix}
    \beta(z+C)\\
    \alpha(z+C)\\
    \gamma(z+C)
    \end{bmatrix}
    =
    \begin{bmatrix}
    c^2 & -2\,c\,s & s^2\\
    -c\,c' & c\, s' + c'\,s & -s\, s'\\
    {c'}^2 & -2\,c'\,s' & {s'}^2
    \end{bmatrix}
    \begin{bmatrix}
    \beta(z)\\
    \alpha(z)\\
    \gamma(z)
    \end{bmatrix} \label{eq:twissturntransportmatrix},
\end{align}
where $C$ is the circumference of the ring and the $c(z)$,  $c'(z)$,  $s(z)$ and $s'(z)$ parameters are those of the turn transport matrix at location $z$. The solution of this eigenvalue problem yields
\begin{align}
    \beta(z) = \frac{2\,s}{\sqrt{(2-c-s')(2+c+s')}} \label{eq:betaeigenvector},
\end{align}
the value of $\beta$ depend thus on the longitudinal position $z$ where the equation is solved. 
The value of $\beta$ obtained with equation \eqref{eq:betaeigenvector} can be used with equation \eqref{eq:betaepsilon} to derive the emittance from a single beam profile measurement
\begin{align}
    \varepsilon=\frac{\sigma^2}{\beta} \label{emittancefromsigma}.
\end{align}
This is the typical method used to measure the beam emittance in a ring. In reality more precise values of $\beta$ are obtained with particle tracking codes that include field errors, misalignment's, non linearities etc. In many cases the beta function can be measured directly using techniques like k-modulation or phase-advance analysis~\cite{Tomas:2010zzb}.

Earlier in the chapter it was suggested to chose a location with zero dispersion for the measurement of the emittance, some times this is however impossible. In case the transverse and longitudinal distributions of the particles are Gaussian, and the dispersion ($D$) and the momentum spread ($\Delta p/p$) are known, it is possible to deconvolute the contributions of the dispersion from the measured beam size.
\begin{align}
    \sigma= \sqrt{\beta\varepsilon + \left( D \frac{\Delta P}{p}\right)^2}    
\end{align}

\subsection{Slice emittance}
In high energy electron linacs all the particles travel at speeds very close to the speed of light so that the particles at the head of the bunch will always remain at the head and particles in the tails will always remain in the tails, there is no synchrotron motion typical of ring accelerators or sub-relativistic linacs. This freezing of the longitudinal plane means the emittance can vary along the bunch due to the different energy, space-charge and weak fields experienced by the particles.

In some applications, like coherent light sources based on free electron lasers (FEL) only a small longitudinal section of the bunch is actually used for lasing and it is important to know the emittance of that particular portion. To this end a time resolved emittance measurement is required. In the mentioned facilities the bunches are very short, less than one picosecond. The only instrument with sufficient resolution is the streak camera, alternatively a deflecting RF cavity can be used to \emph{streak} the bunch directly like in a streak camera tube.

\begin{figure}[htb]
\centering
\includegraphics[align=c, width=0.9\linewidth]{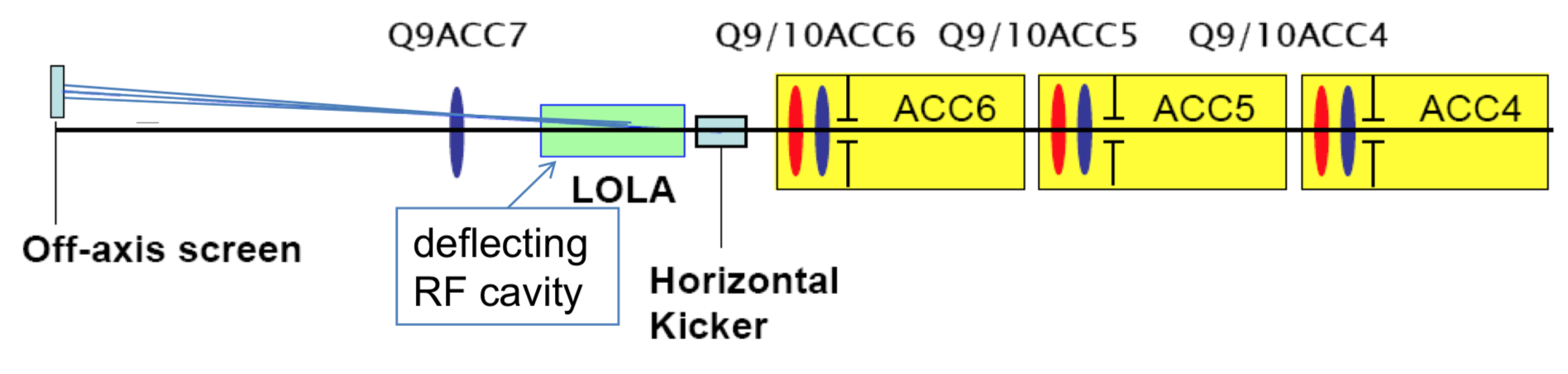}
\caption{Sketch of the slice emittance measurement at FLASH in DESY.}
\label{fig:sliceemittanceflash}
\end{figure}

\Figure[b]~\ref{fig:sliceemittanceflash} shows the setup used in the X-FEL FLASH at DESY. At the end of the linac the beam is deflected onto an off axis screen by a kicker magnet, a deflecting RF cavity (LOLA) kicks the bunch again, but this time the head and the tails of the bunch are kicked in opposite directions. The beam image on the screen downstream can be used to measure the transverse size, in the direction orthogonal to the kick, as function of the deflection (time). A quadrupole right after the deflecting cavity is used to amplify the small deflection and to perform a quad-scan emittance measurement, more information can be found in \Bref{Bolzmann:2005vw}.

\bibliographystyle{unsrt}
\bibliography{transverse_emittance}

\end{document}